\documentclass[conference,a4paper]{IEEEtran}
\usepackage{etex}
\usepackage{cite}
\ifCLASSINFOpdf
\usepackage[pdftex]{graphicx}
\else
\usepackage[dvips]{graphicx}
\fi
\usepackage[cmex10]{amsmath}
\usepackage{algorithmic}
\usepackage{array}
\usepackage{mdwmath}
\usepackage{mdwtab}
\usepackage{eqparbox}
\usepackage[tight,footnotesize]{subfigure}
\usepackage{fixltx2e}
\usepackage{stfloats}
\ifCLASSOPTIONcaptionsoff
\usepackage[nomarkers]{endfloat}
\let\MYoriglatexcaption\caption
\renewcommand{\caption}[2][\relax]{\MYoriglatexcaption[#2]{#2}}
\fi
\usepackage{epsfig}
\usepackage{graphicx}
\usepackage[outdir=./]{epstopdf}

\usepackage{caption}
\usepackage{morefloats}
\usepackage{float}
\usepackage{multirow}
\usepackage{color}
\usepackage{subfigure}
\usepackage{amssymb}
\usepackage{color}
\usepackage{tikz}
\usetikzlibrary{shapes,arrows,decorations.markings}

\usepackage{enumitem}
\setitemize{noitemsep,topsep=0pt,parsep=0pt,partopsep=0pt}
\usepackage[a4paper, top=1.9cm, bottom=4.3cm, left=1.41cm, right=1.41cm]{geometry}

\newcommand\blfootnote[1]{%
	\begingroup
	\renewcommand\thefootnote{}\footnote{#1}%
	\addtocounter{footnote}{-1}%
	\endgroup
}

\begin{document}
\title{Security Trust Zone in 5G Networks}
\author{
Bin~Han\IEEEauthorrefmark{1}, Stan~Wong\IEEEauthorrefmark{2}, Christian~Mannweiler\IEEEauthorrefmark{3}
Mischa~Dohler\IEEEauthorrefmark{2}, Hans~D.~Schotten\IEEEauthorrefmark{1}\\
\IEEEauthorrefmark{1} Technische Universit\"at Kaiserslautern, 67663 Kaiserslautern, Germany \\
\{binhan, schotten\}@eit.uni-kl.de\\
\IEEEauthorrefmark{2}King's College London, Strand, London WC2R 2LS, England, United Kingdom\\
\{Stan.Wong, Mischa.Dohler\}@kcl.ac.uk\\
\IEEEauthorrefmark{3}Nokia Bell Labs, 81541 Munich, Germany\\
christian.mannweiler@nokia-bell-labs.com\\
}

\maketitle

\begin{abstract}
Fifth Generation (5G) telecommunication system is going to deliver a flexible radio access network (RAN). Security functions such as authorization, authentication and accounting (AAA) are expected to be distributed from central clouds to edge clouds. We propose a novel architectural security solution that applies to 5G networks. It is called Trust Zone (TZ) that is designed as an enhancement of the 5G AAA in the edge cloud. TZ also provides an autonomous and decentralized security policy for different tenants under variable network conditions. TZ also initiates an ability of disaster cognition and extends the security functionalities to a set of flexible and highly available emergency services in the edge cloud.
\end{abstract}


\IEEEpeerreviewmaketitle

\section{Introduction}
\blfootnote{This is a preprint, the full paper has been published in the 2017 24th International Conference on Telecommunications (ICT), \copyright 2017 IEEE. Personal use of this material is permitted. However, permission to use this material for any other purposes must be obtained from the IEEE by sending a request to pubs-permissions@ieee.org.}

\IEEEPARstart{F}{ifth} generation (5G) is expected to develop a flexible radio access network (RAN) and to provide the necessary adaptability for handling the fluctuations in the traffic demands. It is also intended to deliver an independent control of logical network slice and to provide an isolatable network resource for the tenants with their plethoric network services (e.g., voice network services, vehicles network services, the Internet of Things network services etc.).

These ambitions have lead to a requirement of high degrees of flexibility and decentralization in the network security functions, such as authentication, authorization and accounting (AAA). Supported by the modern technology of network function virtualization (NFV), a novel hierarchical and distributed AAA approach, the 5G AAA, has been proposed. It combines two independent international standard systems that are the Third Generation Partnership Project (3GPP) and the European Telecommunications Standards Institute (ETSI), as a single platform to manage and secure the subscribers, tenants and network slices under the 5G flexible network environment. By distributing databases to all edge clouds, it enables a flexible and decentralized decision and application of security policies in every edge cloud.

However, concentrating on the vertical dimension of its hierarchical topology, the proposed 5G AAA approach lacks details about its horizontal implementation in the edge clouds. It still remains ambiguous for the edge clouds, how to coordinate the administrations of the hierarchical and distributed AAA servers, or the accesses to the hierarchical and distributed subscriber databases. This issue becomes even more complex, when considering that the connections between edge clouds and central clouds may be restricted under certain conditions, which increases the difficulty of network function management and brings an extra risk of data leakage in.

In this paper, we propose a Trust Zone (TZ) as an edge cloud architectural solution, to  horizontally enrich and extend the 5G AAA approach. With a state model depending on the central cloud availability, TZ is able to adaptively manage its security administration and database accesses. By building an entity model with five functional modules, we seamlessly integrate the TZ design with the 5G AAA architecture. To mitigate security risks during reconnection between central cloud and edge cloud, we propose a safe approach of transferring the access management between central cloud and edge cloud.

This paper is organized as follow: as a background, in Sec. \ref{sec:5g_arch} and Sec. \ref{sec:5g_aaa} we briefly introduce the overall 5G network architecture and the 5G AAA approach, respectively. Subsequently, as the main focus of this work, the Trust Zone design is presented in Sec. \ref{sec:tz}, including its concept, use cases, state model and entity model. Afterwards, the access management transferring approach is discussed in Sec. \ref{sec:secur_mech}, before we close this article with our conclusion in Sec. \ref{sec:conclusion}.

\section{5G Mobile Network Architecture}
\label{sec:5g_arch}

\subsection{The 3GPP NextGen Architecture}
\label{sec:5g_arch_3GPP}
In \cite{3gpp:23.799}, the 3GPP technical specification group on architecture (SA2) currently works on the high-level system architecture as the collection of required capabilities, and high level functions with their interactions between each other. Among others, it proposes the non-roaming ''NextGen Architecture'' consisting of the following high-level functions: authentication server function (AUSF), unified data management (UDM), core access and mobility management function (AMF), session management function (SMF), policy control function (PCF), user plane functions (UPF) and RAN. Further, agreements on the key issue of network slicing indicate that the RAN can be common to multiple network slices. For the core network, SA2 differentiates between common control network functions and dedicated control functions, while the user plane is generally assumed to be dedicated to a specific slice.
\subsection{The 5G PPP Architectures}
\label{sec:5g_ppp}
5G Infrastructure Public Private Partnership (5G PPP) Architecture working group has proposed five different architecture views and a so-called Network Softwarization and Programmability Framework to facilitate the development of the 5G network architecture \cite{5GPPP:arch}. The former focus on the various relevant perspective on 5G system design, which are the application and business service view, infrastructure control view, logical \& functional view, physical resources view, and system management view. The latter introduces a separation into five planes that enable programmable, softwarized networks: Application and Business Service Plane, Multi-Service Management Plane, Integrated Network Management \& Operations Plane, Infrastructure Softwarization Plane, Infrastructure  Control  Plane, and Forwarding (Data) Plane.
However, \cite{5GPPP:arch} lacks detailed considerations for the security domain. In particular, the question of how to realize "Security by Design" needs to be answered. While 5G concepts aim at transferring the flexibility of software-defined and virtualized network functionality to the telecommunication domain, they at the same time increase the risk for experiencing security threats known from IT environments, e.g., threatening confidentiality, data integrity, and network availability, or the possibility of mis-allocation of resources among multiple network tenants. 
\subsection{Security Components in the 5G NORMA Architecture}
\label{sec:5g_ppp}
\begin{figure}[!ht]
\centering
\includegraphics[width=.41\textwidth]{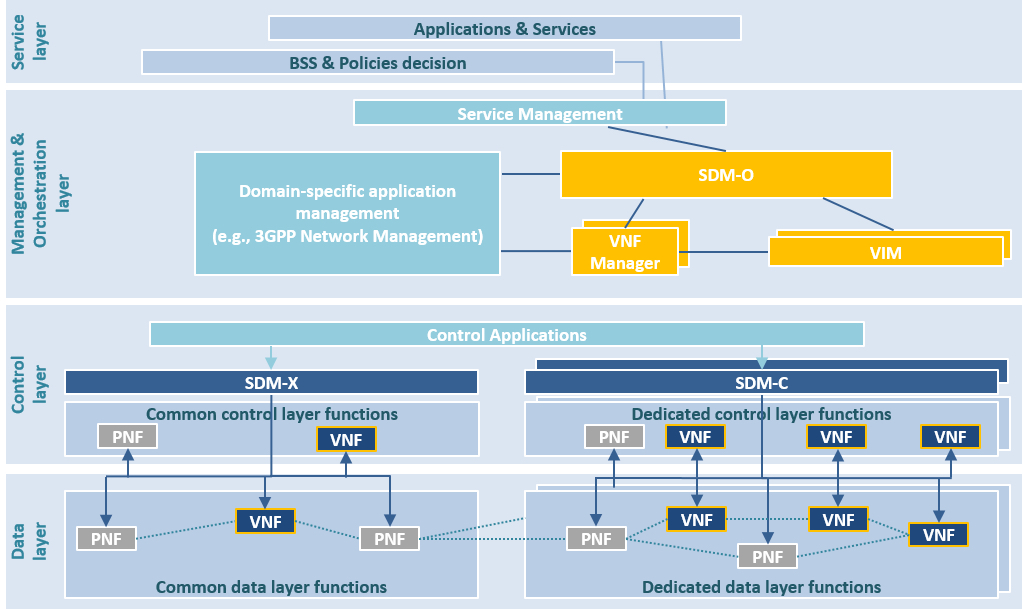}
\caption{High-level 5G NORMA functional architecture}
\label{fig:5gn-arch}
\end{figure}

\begin{figure}[!ht]
\centering
\includegraphics[width=.4\textwidth]{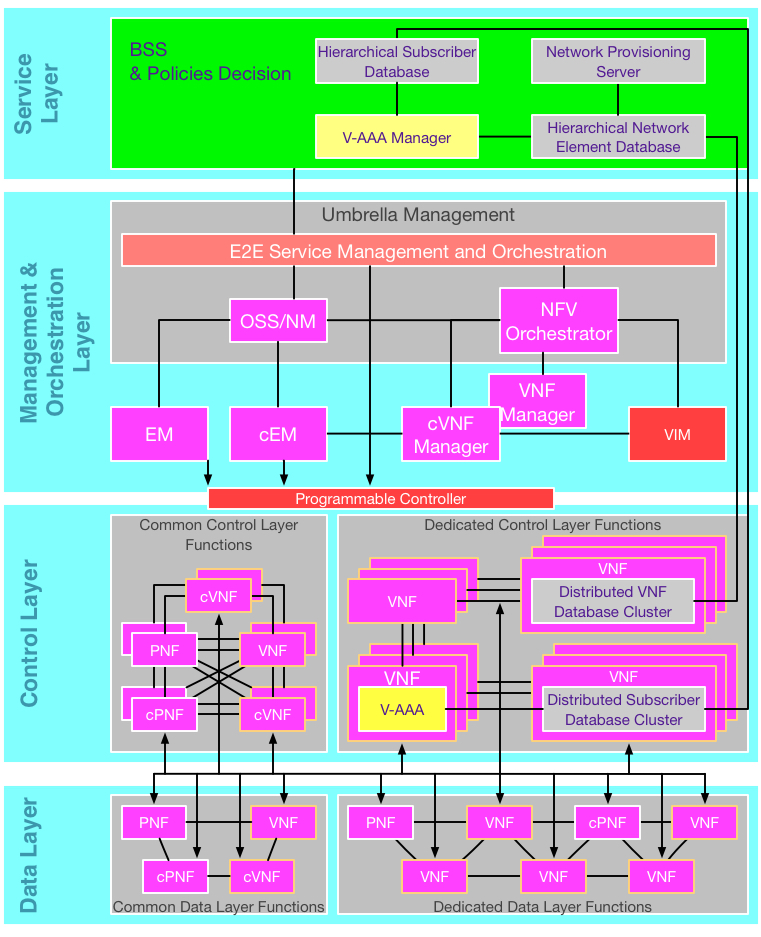}
\caption{5G AAA approach (V-AAA Manager and V-AAA) integrates with 5G PPP Architecture with no tenant and network slice.}
\label{vaaa1}
\end{figure}

Fig.~\ref{fig:5gn-arch} depicts the 5G NORMA overall architecture, comprising four layers: the service layer, management \& orchestration layer, control layer, and data layer \cite{5GN:D32}.
The main design goals for the 5G NORMA architecture comprise adaptive (de-)composition and allocation of network functions, programmable network behavior, as well as joint optimization of network functions. Applying a systematic design approach ensures that the 5G NORMA architecture will securely implement multi-tenant, multi-service networks, building on enabling technologies, such as, multi-level network slicing and software-defined network control and management. More specifically, the authors propose a hierarchical and distributed 5G AAA and Trust Zone concept as an inherent component of the 5G network architecture. 
Fig.~\ref{vaaa1} shows a basic scenario of the 5G PPP logical framework control, management and orchestration of network functions integrated with 5G AAA. Basically, it adopts the ETSI NFV entities, i.e. Virtual Infrastructure Manager (VIM), VNF Manager (VNFM) and NFV Orchestrator (NFVO), and the ONF SDN programmable control layer and data layer. The two layers comprise both  common and dedicated network functions. While common control and data layer functions are shared by multiple network slices, dedicated control and data layer functions are specifically allocated to a single network slice.
In the following sections, we will take a step-by-step approach to explain our approach of the 5G AAA hierarchical and distributed V-AAA, and the details of Trust Zone interfaces and functionalities.

\section{5G\,Authentication-Authorization-Accounting}
\label{sec:5g_aaa}
In this section, we discuss the 5G AAA design bases on the 3GPP AKA, non-3GPP EAP-AKA, and ETSI NFV, reuse of the existing cryptographic functions in 3GPP and tokenization technique in OpenStack KeyStone, and provision of the remote method invocation to other network entities and billing platform for subscribers, tenants and tenant's subscribers. 

The 5G AAA is: (i) to assist tenants isolation and tenant data isolation, (ii) to support tenant's data replication in many-to-one manner from the access network (edge cloud) to core network (central cloud) and local bi-directional replication approaches within tenant's network slice, and (iii) to maintain the central governance in the mobile network operator (MNO) core network and when tenant has been authorized to have a full control of their network slice which tenant can also manage their subscribers in the access network (edge cloud). In this situation, 5G AAA may or may not need to share and move some of the core network functionalities [e.g., access and mobility management function (AMF) or local subscriber server (LSS)] to the access network (edge cloud) which depends on the service level agreement (SLA). It would also provide a better accuracy of UE point of attachment information under the complex of the Next Generation Mobile Networks (NGMN) multi-tenancy, multi-network slicing, multi-level service and multi-connectivity environment \cite{NGMN:arch}. Last but not least, the additional objective is to increase the mobility efficiency of subscribers at the access network (edge cloud) and to reduce the traffic between access network (edge cloud) and core network (central cloud).

Subsequently, the 5G AAA approach converts the traditional macro-management to a micromanagement per-regional based or even per-tenant based. For example, traditionally, MNO applies and uses the 3GPP AKA as an enforcement of the overall security management that remains at the core network, then release the partial right to the access network via different level of cryptographic functions. In contrast, the 5G AAA takes a hierarchical, distributed and dedicated security management approach that can be located within the current LTE eNodeB and only responsible for security management within the eNodeB region. It can also be located within tenant's network slice then the security management responsibility scope is the entire Tenant's network slice. Furthermore, it depends on the MNO and Tenant to configure the scope of security management and to locate the V-AAA entity in the edge cloud. This 5G AAA approach also enhances the flexibility in security management, the accuracy of tracking information i.e. mobility and billing information etc., and the isolation of a tenant’s end-user based on its own geolocation database, which is equivalent to the current LTE eNodeB locations. For example, the first security goal in 3GPP (authentication) is to verify the UE’s identity. While legacy networks perform this verification in the core network, the functionality could be shifted to the edge cloud. A comprehensive explanation of the hierarchical and distributed databases approach for subscribers, tenant and tenant’s subscribers, and under different network integrations of V-AAA into 5G PPP architecture framework are given in section \ref{vaaa1}.
\begin{figure}[!ht]
	\centering
	\includegraphics[width=.45\textwidth]{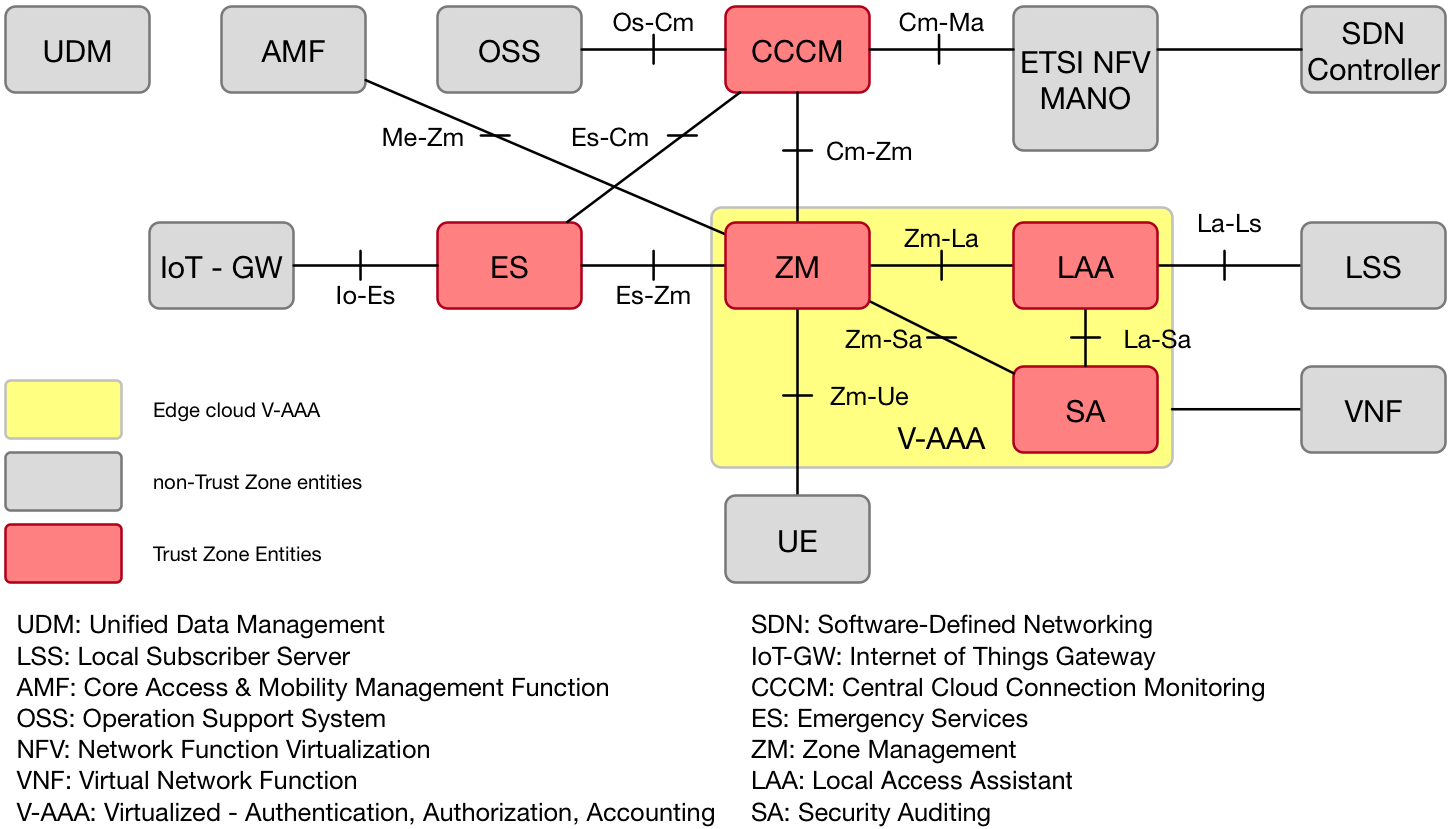}
	\caption{Trust Zone integrated with edge cloud V-AAA server}
    \label{fig:v_aaa_with_tz}
\end{figure}

\section{Trust Zone}\label{sec:tz}

\subsection{Definition of Trust Zone}
In previous section, we discussed the 5G AAA architecture is expected to provide AAA functions flexibly and distributed, in order to support a fully autonomous policy implementation covering the central clouds and the edge clouds. As a connection between edge cloud and central cloud may be - intentionally or unintentionally - restricted, congested or even cut off, the functional availability of edge clouds shall be decoupled from the central cloud to the utmost extent.

Motivated by this challenge, we propose the concept of Trust Zone (TZ), which is defined as a set of network functions covering a geographical area served by a local cell i.e. an edge cloud. In a TZ, different policies are autonomously implemented to ensure data security, while as many services as possible can be provided, regardless of the connection status between this edge cloud and the central cloud. Due to the concern of tenant dependency requirements in AAA services, multiple individual TZs are able to coexist in one edge cloud, each TZ for a different tenant.

The TZ strongly relies on distributed AAA functions, it is tightly integrated with the V-AAA framework. Generally, a TZ is an edge cloud V-AAA server extended with network monitoring function and emergency services.

\subsection{Use Cases}

An architectural solution to distributed security and emergency services, TZ widely plays a role in many 5G communication use cases, where critical communications happen within the local edge cloud, and the intra-edge-cloud connections can remain available under a limited or absent edge-cloud-to-central-cloud connection (EC4). These use cases are including not limited to
industry control, sensor networks monitoring, massive nomadic/mobile machine-type-communication, vehicle-to-anything (V2X) communications, and emergency communications \cite{5GN:D32}. 

\subsection{State Model}
To describe the behavior of TZ, a state model is built. At an arbitrary time instant, according to the status of EC4, a Trust Zone is in one of the following five states: 
Connected (\textit{C}), Weakly Connected (\textit{W}), Lost connection (\textit{L}), Reconnecting (\textit{R}) and Disconnecting (\textit{D}). In the state \textit{C}, the EC4 remains available and healthy as normal. In \textit{W}, the EC4 remains available, but too weak to completely maintain the usual set of network functions that need support of central cloud security functions. When the state is \textit{L}, the EC4 remains unavailable, i.e. no message exchange between the edge cloud and the central cloud can be executed. \textit{R} presents that the EC4 has just been recovered from an usual status. \textit{D} means that the EC4 was in a normal or weak status, but has just vanished.

\begin{figure}[!h]
\centering
\includegraphics[width=.33\textwidth]{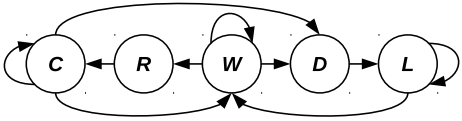}
\caption{The TZ state model. The abbreviations \textit{C}, \textit{R}, \textit{W}, \textit{D} and \textit{L} stand for the states Connected, Reconnecting, Weakly Connected, Disconnecting and Lost Connection, respectively.}
\label{fig:state_model}
\end{figure}

Possible transitions between different states are illustrated in Fig. \ref{fig:state_model}. The states \textit{C}, \textit{W} and \textit{L} are steady states, in which the TZ can remain for a certain duration. \textit{R} and \textit{D} are transient states, which last for only a short time before the TZ state turns into \textit{C} and \textit{L}, respectively. Normally, a TZ remains in its \textit{C} state. When affected by congestions, disasters or attacks, the TZ may turn into its \textit{W} or \textit{L} state. Usually, the EC4 becomes significantly weaker before it completely vanishes, which can be represented by a \textit{C-W-D-L} transition chain. Only in rare cases, a healthy EC4 can be completely and immediately disconnected, which is described by a \textit{C-D-L} transition chain. When the EC4 is recovered, the state turns back through a \textit{W-R-C} or \textit{L-W-R-C} transition chain. We consider an immediate full recovery of EC4 from a complete disconnection as impossible, the \textit{L-R} transition is therefore invalid in our state model.

\subsection{Entity Model}
In a structural view, the functionalities needed to implement a TZ are organized in five functional entities, which have already been illustrated in Fig. \ref{fig:v_aaa_with_tz}. These entities, and the interfaces connecting them, construct the entity model of TZ.

\subsubsection{Central Cloud Connection Monitoring (CCCM)}
According to our state model, the state of a Trust Zone depends on the EC4 status, which is monitored in real time by the CCCM module. Generally, CCCM periodically visits the OSS and the NFV-MANO, to evaluate and predict the EC4 status. Additionally, in the TZ states \textit{W} and \textit{L}, i.e. when the EC4 is limited or disconnected, CCCM tries to diagnose the malfunction. This diagnose can be used by the software-defined network (SDN) management to help recover the EC4 by deploying backup resources, e.g. network redundancy and alternative routes such as satellite links. A dynamic network resource allocation in the SDN management can be also supported by the EC4 information, so that in the TZ state \textit{W}, special network functions such as user authentication/authorization and synchronization of subscriber data can obtain network resources with higher priorities.
\subsubsection{Zone Management (ZM)}
The central controlling entity of TZ, TM is connected with every other TZ entity. It triggers and coordinates the state transition in the entire TZ, when it receives a report of change in the EC4 status. Integrated with interfaces to the central cloud and to the UEs, it also collaborates with the AMF and the LAA to accomplish the Access Stratum (AS) security procedures in the local base station.
When the EC4 is healthy (TZ state \textit{C}), ZM cooperates with the central cloud V-AAA server to provide normal security services. When the EC4 is limited or unavailable (TZ states \textit{W} and \textit{L}), decentralized AAA services are needed, ZM will then collaborate with LAA instead of the central cloud. 
\subsubsection{Local Access Assistant (LAA)}
The LAA is responsible to support a central-cloud-independent local user access procedure when the AMF is unavailable, by performing partial functionalities of the AMF, e.g. deriving keys. 
LAA is deactivated by the ZM when the TZ state is \textit{R}, remains inactive when \textit{C} or \textit{W}, and activated when \textit{D}. In the TZ state \textit{L}, it remains active, exchanging control plane messages with the UEs, deriving/updating the AS keys for them, and providing them with other AS security services.
To enable these functions, the LAA makes use of the user security log files stored in the local subscriber database, which are periodically updated and synchronized with the hierarchical subscriber database in the central cloud. The LAA is isolated from the ZM mainly due to the security considerations, that the subscriber database should be decoupled from the ZM, which is the central controller of TZ and the main target of potential cyber-attacks. Its functionalities are strictly limited in the AS domain, as the NAS keys can only be generated in the central cloud.
\subsubsection{Security Auditing (SA)}
MNOs usually keeps a log of security critical operations for each user, these logs can be audited to investigate all potential risks of illegal access and cyber-attacks. Both the log database and the auditing centre are usually located in the central cloud, so that they cannot keep tracking the authentication and authorization operations locally executed by the modules ZM and LAA in the TZ state \textit{L}. To close this gap and promise a seamless audit, the SA module is implemented. It is activated by the ZM in the TZ state D, recording all security critical operations in the state \textit{L}. These records can be either actively pushed to the central auditing center in the state \textit{R}, or passively pulled in the state \textit{C} upon need.
\subsubsection{Emergency Services (ES)}
Considering the high capacity and robustness of 5G backhaul networks, the appearance of EC4 quality damage usually implies emergency situations, which can cause a massive physical damage on the network infrastructures and an impulsive peak of service demand, e.g. earthquakes, fires, explosions, etc. A set of emergency services can be defined, which help users avoid personal injury and property damage under such disasters, even when their devices cannot be authenticated or authorized. These services include but are not limited to
public disaster alarm, evacuation guidance, positioning service, emergency call and short message service (SMS).

Some functions on this list shall be only valid under specific disasters, e.g. public disaster alarm and evacuation guidance. Some others, such like SMS, shall be always available, but provided with different security policies under different situations. The rest, such as emergency call, shall always remain available without authentication and authorization, regardless of the situation. To achieve this, the EM module receives the TZ state information from the ZM, and collects disaster information from public security infrastructures via the Internet-of-things (IoT). According to the information, it autonomously makes the security policy decision for each individual emergency service. It also forwards the disaster information to the CCCM to help diagnose the EC4 malfunctions.

\subsubsection{Interfaces}
To manage the exchange of data, messages and commands among Trust Zone, V-AAA and other 5G NORMA functional entities, a set of interfaces are defined, as illustrated in Fig. \ref{fig:v_aaa_with_tz}:

\begin{itemize}
\item\textit{Cm-Ma} enables CCCM to visit NFV-MANO for EC4 evaluation. When the EC4 is in error, CCCM also sends the diagnose to NFV-MANO through it.
\item\textit{Cm-Zm} enables CCCM to report ZM about the EC4 status and to get state transition messages from ZM.
\item\textit{Es-Cm} forwards disaster alarms from ES to CCCM.
\item\textit{Es-Zm} enables ZM to trigger TZ state transitions in ES.
\item\textit{Io-Es} delivers disaster alarms from IoT to ES.
\item\textit{La-Ls} delivers the synchronized local copies of user profile from LSS to LAA.
\item\textit{La-Sa} enables SA to monitor the operations of LAA.
\item\textit{Me-Zm} allows ZM to receive keys derived by AMF.
\item\textit{Os-Cm} enables CCCM to visit OSS for EC4 evaluation.
\item\textit{Zm-La} delivers edge-cloud-derived keys from LAA to ZM, and TZ state transition messages from ZM to LAA.
\item\textit{Zm-Sa} enables SA to monitor the operations of ZM.
\item\textit{Zm-Ue} allows ZM and UEs to exchange C-plane messages for the security procedure.
\end{itemize}

\subsection{Integration with V-AAA}
To integrate TZ with the aforementioned 5G AAA approach, we first investigate the location of  TZ entities. The functionalities of ZM, LAA, SA and EM are limited in edge cloud. They can be locally implemented. In contrast, CCCM must be implemented over edge and central clouds, to realize the EC4 evaluation. Then we consider their role in the 5G AAA approach. The basic functionality set essential for local AAA is covered by the entities ZM, LAA and SA, while CCCM and EM only extend them with environment cognition and special services. Hence, ZM, LAA and SA together constitute the local V-AAA server, as shown in Fig. \ref{fig:v_aaa_with_tz}

\section{Safe Approach to Transfer Access Management}\label{sec:secur_mech}

Compared to the V-AAA manager, local V-AAA servers are less secured due to the incomplete set of security functionalities. Additionally, considering the large total number of the local V-AAA servers, they also probably have less budget on security measures. This brings a risk to the idea of distributing security functions at the edge cloud. Social engineers may initiate attacks to disconnect the central cloud and the edge cloud, and may hack the local TZ, which is easier to target than the central cloud. Then they might eventually try to obtain access to the central cloud during the reconnection, when the edge cloud hands its security functions back over to the central cloud. To avoid this risk, an asymmetric approach of transferring the access management between the central cloud and the trust zone is designed as follows.

When a disconnection takes place (state \textit{D}), the ZM considers all UEs that have already been authenticated as trusted devices. These devices are able to retain maximal access to the TZ according to the respective policy under the current situation, until they lose their connections to the edge cloud.

When a UE tries to access the TZ and the central cloud is unavailable (state \textit{L}), the ZM invocates the LAA and the LSS to gain an access for the UE. If the subscriber data of the UE is available in LSS and the security check is passed, the UE can be considered as a trusted device. Otherwise, it remains untrusted and is only granted the basic emergency services.

When the edge cloud is reconnected to the central cloud (state \textit{R}), the ZM disconnects all UEs in the local TZ in a pre-scheduled order, so that the UEs have to be re-authenticated and reauthorized by the central security server, in order to regain full access.

With this mechanism, emergency services are ensured to remain available for all users, edge cloud services are as much attainable as possible for legal users, while fake devices are prevented from accessing the central cloud.

\section{Conclusion}\label{sec:conclusion}
In this paper, we provided a condense study of 5G PPP logical architecture and the 5G potential AAA approach. Particularly, we have proposed a novel approach Trust Zone (TZ) as an edge cloud solution of distributed 5G security. Through an EC4 quality estimation mechanism, the TZ method cognitively enables the security functions in central or edge clouds, and seamlessly integrated with the distributed V-AAA in the edge cloud. By defining emergency services and supporting external services e.g., IoT services, TZ improves the security of 5G access network (edge cloud) in two aspects: the flexibility of security management and the resistance to disasters. The security risk is reduced by an asymmetric approach of transferring the access management. Besides the security issues discussed in this paper, the TZ also exhibits potential of shifting other central cloud functionalities into edge cloud, which could increase the network flexibility. For future works, studies alike \cite{bellavista:virtual} on the implementation efficiency of TZ on different edge computing platforms are expected.
\appendices

\section*{Acknowledgement}
This work has been performed in the framework of the H2020-ICT-2014-2 project 5G NORMA. The authors would like to acknowledge the contributions of their colleagues. This information reflects the consortium’s view, but the consortium is not liable for any use that may be made of any of the information contained therein.

\ifCLASSOPTIONcaptionsoff
  \newpage
\fi


\end{document}